\documentclass[
 reprint,
 amsmath,amssymb,
pra,
]{revtex4}
\usepackage[section]{placeins}
\usepackage{graphicx}% Include figure files
\usepackage{dcolumn}% Align table columns on decimal point
\usepackage{bm}% bold math
\usepackage{xcolor}%to define textcolor
\begin{document}

%\preprint{APS/123-QED}
\title{On Sylvester solution for degenerate eigenvalues}% Force line breaks with \\

\author{Dawit Hiluf }
\email{dawit.hailu@mail.huji.ac.il}

\affiliation{Physics Department, Mekelle University, P.O.Box 231, Mekelle, Ethiopia.}
\date{\today}

\begin{abstract}
In this paper we introduce the use of Sylvester's formula for systems with degenerate eigenvalues in relation to obtaining their analytical solutions. To appreciate the use we include two other forms of analytical solutions namely adiabatic and Magnus approximations. In quantum mechanics, the Schr\"{o}dinger equation is a mathematical equation that describes the evolution over time of a physical system in which quantum effects, such as wave--particle duality, are significant. The equation is a mathematical formulation for studying quantum mechanical systems. Just like Newtons's laws govern the motion of objects, Schr\"{o}dinger equations of motion also govern the motion of quantum objects. Unlike the classical motion of objects the equation of motions of quantum phenomenon deals with the likelihood of the trajectories.
\end{abstract}

\maketitle
%%%%%%%%%%%%%%%%%%%%%%%%%%%%%%%%%%%%%%%%%%%%%%%%%%%%%%%%%%%%%%%%%%%%%%%%%%%%%%%%
%\section{Abstract}
%%%%%%%%%%%%%%%%%%%%%%%%%%%%%%%%%%%%%%%%%%%%%%%%%%%%%%%%%%%%%%%%%%%%%%%%%%%%%%%%
%In this paper we introduce the use of Sylvester's formula for systems with degenerate eigenvalues in relation to obtaining their analytical solutions. To appreciate the use we include two other forms of analytical solutions namely adiabatic and Magnus approximations. In quantum mechanics, the Schr\"{o}dinger equation is a mathematical equation that describes the evolution over time of a physical system in which quantum effects, such as wave--particle duality, are significant. The equation is a mathematical formulation for studying quantum mechanical systems. Just like Newtons's laws govern the motion of objects, Schr\"{o}dinger equations of motion also govern the motion of quantum objects. Unlike the classical motion of objects the equation of motions of quantum phenomenon deals with the likelihood of the trajectories.
%%%%%%%%%%%%%%%%%%%%%%%%%%%%%%%%%%%%%%%%%%%%%%%%%%%%%%%%%%%%%%%%%%%%%%%%%%%%%%%%
\section{Introduction}
%%%%%%%%%%%%%%%%%%%%%%%%%%%%%%%%%%%%%%%%%%%%%%%%%%%%%%%%%%%%%%%%%%%%%%%%%%%%%%%%
Generally speaking physical systems are dynamical, implying that they evolve in time. Such dynamical evolution of quantum mechanical systems are described by Schr\"{o}dinger equation of motion by describe the dynamical properties of quantum systems. Therefore given a state $|\psi(0)\rangle$ of a quantum system at some initial time $t = 0$, what is required is a physical law that tells us what the state will be at some other time $t$, i.e. a quantum law of evolution. Given that a state vector is a repository of the information known about a system, what is required is a general physical law that tells us how this information evolves in time in response to the particular physical circumstances that the system of interest finds itself in. What is being explored is the reaction of the physical system to an external perturbation. It is to be expected that the details of this law will vary from system to system, but it turns out that the law of evolution can be written in a way that holds true for all physical systems. 

Dynamics of quantum mechanical systems that lacks external perturbation can readily be studied using Schr\"{o}dinger equation of motion in which the Hamiltonian of the system is taken to be constant. Such system's solutions sometimes, if not often, are readily provided using analytical approach. Although we must state that doing so requires making assumptions, or approximations, about the system at hand. Such approximations and/or assumptions help us simplify the complexity of the physical system thereby equipping us with the ability to solve the equation of motion analytically. Besides reducing the complexity of the system, these assumptions also help us zoom into the system we are interested in by avoiding the excess information surrounding the physical system. Furthermore the idea of obtaining such solutions is helpful in filtering out the controlling parameters, which not only is useful in proposing experimental designs, but also is crucial in gaining insights on the physical system under investigation. Following Bohr such physical state is commonly known as stationary state. It is a term that is usually used to identify those states of a quantum system that do not change in time. Yet we must underline that, despite the name, such state is really imbued with interesting physics. 

Our aim in this paper is to show use of Sylvester's formula for dynamical systems with degenerate eigenvalues, we will first introduce some approximations and assumptions that justifies the use of analytical solutions. Amongst several methods of analytical solutions, we chose Sylvester theorem as a tool for analytical solution because our interest lies in obtaining analytical solution for quantum system having degenerate eigenvalues. For we are interested in obtaining an analytical solution for a physical system that have degenerate eigenvalues, Sylvester's formula seems a good candidate as it requires only knowledge of eigenvalues. 

Our choice of obtaining solutions via eigenvalue emanates from quantum mechanics. Recall in quantum mechanics, a quantum state can be thought of as a vector, say, in Hilbert space. Moreover measurement on a state is described as operation in a state or time evolution, such as, due to interacting laser light. This in turn can be thought of as a matrix acting on state vector while the eigenvalue represents the energy of the state. Such description holds true for any observable, i.e physically measurable quantity. In quantum mechanics, it is the eigenvalues of these observables that correspond to the actually measured values. 
%%%%%%%%%%%%%%%%%%%%%%%%%%%%%%%%%%%%%%%%%%%%%%%%%%%%%%%%%%%%%%%%%%%%%%%%%%%%%%%%
                                                                             % Chapter one % section one
%%%%%%%%%%%%%%%%%%%%%%%%%%%%%%%%%%%%%%%%%%%%%%%%%%%%%%%%%%%%%%%%%%%%%%%%%%%%%%%%
Strictly speaking observations in quantum physics do not generally have observables as their outcomes but expectation value of the observables. This signifies that the connection between theory and experiment is via the observables' expectation values. It is worth mentioning here that these expectation values are real numbers corresponding to the outcomes of the measurement of the observables. We work in Heisenberg picture, as in this picture, observables are functions of time meaning that they specify the changing expectation values of observables. Mathematically put, the expectation values are given by a fixed linear function,$\langle O\rangle$, from observables to real numbers. 

To reiterate, in quantum mechanics the Schr\"{o}dinger equation is a mathematical equation that describes the evolution over time of a physical system in which quantum effects are significant. The Schr\"{o}dinger equation of motion is a mathematical formulation for studying quantum mechanical systems. This approach requires the knowledge of the Hamiltonian and the initial state of the system at $t=t_0$ to yield the state of the quantum system after some time $t$. The Hamiltonian is the operator corresponding to the total energy of the system in most of the cases. It is usually denoted by $H$, also $\hat H$. Its spectrum is the set of possible outcomes when one measures the total energy of a system. Because of its close relation to the time-evolution of a system, it is of fundamental importance in most formulations of quantum theory. The Hamiltonian generates the time evolution of quantum states.

%Associated with each measurable parameter in a physical system is a quantum mechanical operator, and the operator associated with the system energy is called the Hamiltonian. The Hamiltonian contains the operations associated with the kinetic and potential energies. Operating on the wavefunction with the Hamiltonian produces the Schr\"{o}dinger equation. In the time independent Schr\"{o}dinger equation, the operation may produce specific values for the energy called energy eigenvalues. In addition to its role in determining system energies, the Hamiltonian operator generates the time evolution of the wave function

In this paper we will exploit the Hamiltonian of a system interacting with an electric field. We then proceed to use the Hamiltonian to describe the equation of motion for the system at hand. We can describe the evolution of the system either by using the probability amplitude, time dependent Schr\"{o}dinger equation, or by the using the density matrix formalism, in Liouville description.  Using the equations for the density matrix elements, we will form a set of observables by linear combination of the coherences and populations. We next obtain equation of motion in larger space. We at the end seek analytical solution assuming the perturbation fields to have same time dependence but possibly different strength.
%%%%%%%%%%%%%%%%%%%%%%%%%%%%%%%%%%%%%%%%%%%%%%%%%%%%%%%%%%%%%%%%%%%%%%%%%%%%%%%%
\section{Analytical Approximations}
%%%%%%%%%%%%%%%%%%%%%%%%%%%%%%%%%%%%%%%%%%%%%%%%%%%%%%%%%%%%%%%%%%%%%%%%%%%%%%%%
Once the equations of motion for quantum systems are established, it is obvious that one can solve them numerically. But for the purpose of this paper our interest lies in some of the ways one can tackle analytically. It is to be understood that these solutions are not as exact as obtained through numerical approach, and consequently they are considered as an approximation. Yet often than not they are helpful in gaining some physical insights.  The systems of equations that we aim to tackle in this paper are coupled equation of motions, see \cite{alhassid1977entropy,HIOE:1981aa,hiluf2016link} for its full derivations, and has the form given by the following equation of motion (see eq.\eqref{eqmn4g}), for the observable vector $\vec S$.  The observable vector $\vec S$ comprises of the expectation values of the observables, i.e linear combination of the coherences and populations. As the observable vector lives in larger space it consists of more elements, this means for a quantum system with $N$ distinct states the vector has $N^2-1$ elements if normalization is imposed. For instance in a two level system the observable vector has 3 elements and the vector is commonly referred as Bloch vector\cite{hailu2019su2,DHCNOT}. Key ingredient for the derivation is that the Hamiltonian, and density matrix, are expressible as a linear combination of generators thereby form closed Lie algebra. We note that the discussion given in this paper is equally applicable to time dependent Schr\"{o}dinger equations in Hilbert space too, or other equations of motion with similar form.
\begin{equation}
\begin{aligned}
\frac{d}{dt}\vec{S}=&g\vec S
\end{aligned}
\label{eqmn4g}
\end{equation}
In contrast to Schr\"{o}dinger equation vector $\vec{S}$, in eq.\eqref{eqmn4g}, describes the state of the system while matrix $g$ is its Hamiltonian with dimensions $(N^2-1)\times (N^2-1)$ .
Such equations, as pointed out earlier, can be solved numerically to get their exact solutions. Sometimes, however, it is required to make compromise and seek approximate solutions. This kind of approach yields insights on the physical system at the expense of exactness. We would like to stress, however, that by making adjustments on all or selected parameters we can get closer to the exact solution. Moreover it is also possible, and at times advisable, to make use of analytical calculations to filter out the most essential parameters, or control parameters. This is so because knowing controlling parameters leads to proposing and designing of interesting experimental setups. 

In this section, before diving into exploiting Sylvester's formula to get analytical solutions, we first have a look on two other ways of obtaining analytical solution.  To this end we first briefly look at these two methods, adiabatic and Magnus approximations. 
%%%%%%%%%%%%%%%%%%%%%%%%%%%%%%%%%%%%%%%%%%%%%%%%%%%%%%%%%%%%%%%%%%%%%%%%%%%%%%%%
\subsection{Adiabatic Approximations}
\label{adiab}
%%%%%%%%%%%%%%%%%%%%%%%%%%%%%%%%%%%%%%%%%%%%%%%%%%%%%%%%%%%%%%%%%%%%%%%%%%%%%%%%
In what follows adiabaticity, or adiabatic approximation, in time refers to the very slow change of perturbation on physical system that enables the system to be able to align with the same eigenstate before and after the interaction with the perturbation. This is possible because the slow variation in perturbation allows the system to have ample time to adjust to the instantaneous eigenstate. Loosely speaking this means once prepared in an instantaneous eigenstate the system remains in this state provided that its eigenvalue is separated from the nearest states by a finite energy gap. Physically this is translated to mean that there is no transition between the adiabatic states. This in turn is reflected in the structure of the adiabatic Hamiltonian. The matrix elements of the transformed Hamiltonian, the Hamiltonian in the adiabatic picture, are zero or negligibly small,  except on its diagonal part.  Mathematically this can be achieved by considering a unitary transformation matrix $U$ that diagonalizes the matrix $g$ in eq.\eqref{eqmn4g}. 

Following transformation rules let us now multiply both sides of equation eq.\eqref{eqmn4g} by a unitary transformation matrix $U^{-1}$ whose time derivative is zero
\begin{subequations}
\begin{align}
{U}^{-1}\frac{d}{dt}\vec{S}=&{U}^{-1}g\vec S\\
\frac{d}{dt}\big(U^{-1}\vec S\big)=&U^{-1}g\vec S\label{eqmn4gtr}
\end{align}
\end{subequations}
Upon multiplying RHS of eq.\eqref{eqmn4gtr} by an identity matrix $I=UU^{-1}$ one can readily be able to diagonalize matrix $g$. This transformation yields a diagonal matrix $\Lambda=U^{-1}g U$, whose matrix elements comprises of eigenvalues of $g$.
\begin{equation}
\begin{aligned}
%\frac{d}{dt}\big(U^{-1}\vec S\big)=&U^{-1}g U U^{-1}\vec S\\
\frac{d}{dt}\big(U^{-1}\vec S\big)=&\Lambda \big(U^{-1}\vec S\big)
\end{aligned}
\end{equation}
 Let us, for convenience, denote now the transformed observable vector $\vec S'=U^{-1}\vec S$ therefore the equation of motion for this new observable vector is 
\begin{equation}
\begin{aligned}
\frac{d}{dt}\vec{S'}=&\Lambda\vec S'
\end{aligned}
\end{equation}
The solution of which, recalling we are making adiabatic approximation, takes the form
\begin{equation}
\begin{aligned}
\vec{S'}\left(t\right)=&e^{\Lambda t}\vec S' \left(0\right)
\end{aligned}
\label{eqmn4g2}
\end{equation}
As we are looking the solution for eq.\eqref{eqmn4g}, we can retrieve it from the solution eq.\eqref{eqmn4g2} by multiplying the last equation with $U$ to get back to the original observable vector $\vec S$ as
\begin{subequations}
\begin{align}
U\vec{S'}\left(t\right)=&U e^{\Lambda t}\vec S' \left(0\right)\label{eqsoltr}\\
\vec{S}\left(t\right)=&  U e^{\Lambda t}  U^{-1} \vec S \left(0\right)\label{eqsol}
\end{align}
\end{subequations}
where we have inserted identity on the RHS of eq.\eqref{eqsoltr} to arrive at eq.\eqref{eqsol}. This solution can be expressed component-wise as follows
\begin{equation}
\begin{aligned}
S_k\left(t\right)=&\sum_j  U_{kj} e^{\lambda_j t}  U_{jk}^{-1} \vec S_j \left(0\right)
\end{aligned}
\end{equation}
where $\lambda_j $ is the eigenvalue.

\textbf{Example - i}: Let us now exemplify by applying the adiabatic approximation method for two level system, and obtain the solution of the coherence vector, or Bloch vector.  We here consider a two level system, with a ground $|0\rangle$ and excited $|1\rangle$ states, coupled by a coherent laser light of Rabi frequency $\hbar\Omega$ with detuning $\Delta=\hbar\omega_1-\hbar\omega_0$. Assume also that the system is initially prepared to be on the ground state\\
\textbf{Solution}: First thing we do is construct the coherence vector with generators that are closed under lie algebra as outlined in \cite{hailu2019su2,DHCNOT} and the reference there in. For further use we rewrite here the relation between the generators and the components of the coherence vectors as 
\begin{equation}
\begin{aligned}
u_{01} =& |0\rangle\langle1| + |0\rangle\langle1|,\\
v_{01}=&-i(|0\rangle\langle1| - |0\rangle\langle1|),\\
w_1=&|0\rangle\langle0| - |1\rangle\langle1|
\label{defuvw}
\end{aligned}  
\end{equation}  
We have already seen that the equation of motion for two level system in terms of the $SU(2)$  to be 
\begin{equation}
\begin{aligned}
\frac{d}{dt}\begin{pmatrix}
S_1\\
S_2\\
S_3
\end{pmatrix}=&
 \begin{pmatrix}
  0 & \Delta  & 0 \\
  -\Delta & 0 & -\Omega \\
  0 & \Omega & 0
 \end{pmatrix}
\begin{pmatrix}
S_1\\
S_2\\
S_3
\end{pmatrix}\\
\frac{d}{dt}\vec{S}=&g\vec S
\end{aligned}
\end{equation}
Using Mathematica or Matlab one can find the eigensystems, eigenvalues and eigenvectors respectively, of $g$ to be 
\begin{equation}
\begin{aligned}
\Lambda=&\begin{pmatrix}
0 & 0 & 0\\
0 & -\sqrt{-\Delta^2-\Omega^2} & 0\\
0 & 0 & \sqrt{-\Delta^2-\Omega^2}
\end{pmatrix}\\
U=&
 \begin{pmatrix}
  -\frac{\Omega}{\Delta} & \frac{\Delta}{\Omega}  & \frac{\Delta}{\Omega} \\[0.3em]
  0 & -\frac{\sqrt{-\Delta^2-\Omega^2}}{\Omega} & \frac{\sqrt{-\Delta^2-\Omega^2}}{\Omega} \\[0.3em]
  0 & \Omega & 0
 \end{pmatrix}
\end{aligned}
\end{equation}
Therefore plugging the values and noting that $\vec S\left(0\right)=(0, 0, 1)^T$, as we have prepared the system to be initially on the ground state, which follows from  of $\vec S=(u_{01},v_{01},w_1)^T$, and doing some algebra we find the following result
\begin{equation}
\begin{aligned}
\vec{S}\left(t\right)=&  U e^{\Lambda t}  U^{-1} \vec S \left(0\right)\\
\vec{S}\left(t\right)=-&
 \begin{pmatrix}
  \Big(-1+\frac{1}{2}\big(e^{\lambda_1 t}+e^{\lambda_2 t}\big)\Big)\frac{\Delta\Omega}{\Delta^2+\Omega^2}\\[0.6em]
  \frac{1}{2}\big(-e^{\lambda_1 t}+e^{\lambda_2 t}\big)\frac{\Omega\sqrt{-\Delta^2-\Omega^2}}{\Delta^2+\Omega^2}\\[0.6em]
  \Big(\Delta^2+\frac{\Omega^2}{2}\big(e^{\lambda_1 t}+e^{\lambda_2 t}\big)\Big)\frac{1}{\Delta^2+\Omega^2}
 \end{pmatrix}
\end{aligned}
\end{equation}
If we introduce $\Omega_0=\sqrt{-\Delta^2-\Omega^2}$ which yields that $\lambda_1=-\sqrt{-\Delta^2-\Omega^2}=-\imath\Omega_0$, in like manner we can also obtain that $\lambda_2=\sqrt{-\Delta^2-\Omega^2}=\imath\Omega_0$, with the aid of which the solutions can be rewritten as 
\begin{equation}
\begin{aligned}
\vec{S}\left(t\right)=&
 \begin{pmatrix}
  \frac{\Delta\Omega}{\Omega^2_0}\big(1-\cos\Omega_0 t\big)\\[0.6em]
  \frac{\Omega}{\Omega_0}\sin\Omega_0 t\\[0.6em]
 -\frac{\Delta^2}{\Omega^2_0}-\frac{\Omega^2}{\Omega^2_0}\cos\Omega_0 t
 \end{pmatrix}
\end{aligned}
\label{soln2lvl}
\end{equation}
To exploit the benefit of having analytical solution, we now take a pause and see what we can learn from the solution just obtained, i.e. eq.\eqref{soln2lvl}. To begin with from the definition we used to construct the coherence vector, see eq.\eqref{defuvw}, it is worth pointing out that the solution corresponds to evolution of the vector at time $t$ after the system is perturbed by a laser field. Where elements of the coherence vector comprises of real and imaginary parts of the coherences (i.e. $\rho_{01}$ of the density matrix), and the population differences between the two levels (i.e. $\rho_{aa}, a=0,1$). Furthermore we use the solution as a spring board for what we want to achieve. For instance, to use the most commonly known example of complete population transfer in two level state system our target would have been to obtain $\vec S\left(t\right)=(0, 0, -1)^T$. Looking at eq.\eqref{soln2lvl} we require that  
\begin{equation}
\begin{aligned}
  \frac{\Delta\Omega}{\Omega^2_0}\big(1-\cos\Omega_0 t\big)=0\\
  \frac{\Omega}{\Omega_0}\sin\Omega_0 t=0\\
 -\frac{\Delta^2}{\Omega^2_0}-\frac{\Omega^2}{\Omega^2_0}\cos\Omega_0 t = -1
\end{aligned}
\label{soln2lvlana}
\end{equation}
One readily notices that the solution is dependent on $\Delta$ and $\Omega$ along with their combinations, therefore from these simultaneous equations we learn that the desired condition would be fulfilled when $\Omega_0 t = \pm\pi$  for negligibly small detuning $\Delta\rightarrow0$, which thus consequently implies $\Omega t = \pm i\pi$. More specifically the solution on resonance, where $\Delta=0$, takes the form 
\begin{equation}
\begin{aligned}
\vec{S}\left(t\right)=-&
 \begin{pmatrix}
  0\\
  \frac{\imath}{2}\big(e^{-\imath\Omega t}-e^{\imath\Omega t}\big)\\[0.6em]
 -\frac{1}{2}\big(e^{-\imath\Omega t}+e^{\imath\Omega t}\big)
 \end{pmatrix}
\end{aligned}
\end{equation}
making use of $\lambda_1=-\sqrt{-\Delta^2-\Omega^2}=-\imath\Omega_0$ and $\lambda_2=\sqrt{-\Delta^2-\Omega^2}=\imath\Omega_0$ it becomes
\begin{equation}
\begin{aligned}
\vec{S}\left(t\right)=&
 \begin{pmatrix}
  0\\
 \sin\Omega_0 t\\
 -\cos\Omega_0 t
 \end{pmatrix}
\end{aligned}
\end{equation}
%\textbf{Check if it is $\Omega_0$ or $\Omega$}
If on the other hand the desire is to create coherences between the ground and the excited states, say by distributing populations equally amongst the two states, what we aspire is to achieve is either $\vec S\left(t\right)=(\frac{1}{2}, 0, 0)^T$ or $\vec S\left(t\right)=(0, \frac{1}{2}, 0)^T$ depending on non-zero real or imaginary, respectively, part of the off-diagonal element in the density matrix.  Solution for for eq.\eqref{soln2lvl} yields several conditions to be met, amongst which  $\Omega t = \pm\frac{\pi}{2}$ with $\Delta t=\pm\frac{\pi}{2}$ are conditions to be met.

Following same footsteps, as above, one can obtain several other solutions depending on the desired population distributions such as transferring third or fourth etc of the populations to the excited state or vice versa; this solution in turn can be used by an experimentalist to design and execute an experiment that met the condition. The condition $|\Omega t|$ is commonly known as pulse area and is indicative of the strength of the laser pulse.  For example if the pulse area is $\pi$ it means strong laser field and if $\frac{\pi}{2}$ the applied laser field is weak. To reiterate It is in this sense that we say analytical approach is insightful in the understanding of the physics of the system at hand. 
%%%%%%%%%%%%%%%%%%%%%%%%%%%%%%%%%%%%%%%%%%%%%%%%%%%%%%%%%%%%%%%%%%%%%%%%%%%%%%%%
\subsection{Sylvester's Formula}
%%%%%%%%%%%%%%%%%%%%%%%%%%%%%%%%%%%%%%%%%%%%%%%%%%%%%%%%%%%%%%%%%%%%%%%%%%%%%%%%
The solution discussed in subsection (\ref{adiab}) assumes that the Hamiltonian of the system commutes with itself at different times.  But generally speaking the solution to ODEs with time-dependent coefficients such as the one we are dealing here:
\begin{equation}
\begin{aligned}
\frac{d }{dt}\vec{S}\left(t\right)=&g\left(t\right)\vec S \left(t\right), &     \vec S \left(0\right) =S_0
\end{aligned}
\end{equation}
with $\vec{S}\left(t\right)=(S_1\left(t\right)\ldots,S_n\left(t\right))$ and $g\left(t\right)$ is an $n\times n$ matrix, (the general solution) is given by
\begin{equation}
\begin{aligned}
\vec{S}\left(t\right)=& \mathcal{T}\{e^{\int_0^t g\left(t'\right) dt'}\}\vec S \left(0\right), 
\end{aligned}
\end{equation}
where $ \mathcal{T}$ denotes time-ordering,
\begin{equation}
\begin{aligned}
\mathcal{T}\{e^{\int_0^t g\left(t'\right) dt'}\}\equiv & \sum_{n=0}^\infty\frac{1}{n!}\int_0^t\ldots\int_0^t\mathcal{T}\{g\left(t_1'\right)\ldots g\left(t_n'\right)\}\\
=&\sum_{n=0}^\infty\int_0^t dt'\ldots\int_0^{t'_{n-1}}dt'_n~ g\left(t_1'\right)\ldots g\left(t_n'\right)
\end{aligned}
\end{equation}

Assuming the matrices commute at different times, meaning $[g(t_1),g(t_2)]=0, \forall t_1,t_2$, the time-ordered expression takes the form we $e^{\int_0^t g\left(t'\right) dt'}$. Evaluating such exponentials is very well studied \cite{wilcox1967exponential,wei1963lie,wei1964global,magnus1954exponential,Suzuki1976} and beyond the scope of the present paper. For instance the solution for the dynamics of the two level system explored in this paper is attempted by a Magnus series \cite{magnus1954exponential,hailu2019su2,DHCNOT}, yet one can approach the problem, with one (or more) of the various proposals or methods listed in the reference, to approximate or exactly solve them. In this section, however, we use the Sylvester's formula and obtain analytical solution to coupled differential equations
%%%%%%%%%%%%%%%%%%%%%%%%%%%%%%%%%%%%%%%%%%%%%%%%%%%%%%%%%%%%%%%%%%%%%%%%%%%%%%%%
\subsubsection{Sylvester's formula for distinct eigenvalues}
%%%%%%%%%%%%%%%%%%%%%%%%%%%%%%%%%%%%%%%%%%%%%%%%%%%%%%%%%%%%%%%%%%%%%%%%%%%%%%%%
Another way of solving eq.\eqref{eqmn4g} is possible using the very well known Sylvester's formula.The Sylvester's formula provides a tool for obtaining solution to an exponential equations. To this end, given an $N\times N$ coefficient matrix $g\left(t\right)$ we wanted to solve the initial value value problem associated with the linear ordinary differential equation, which in our case is the equation of motion for the coherence vectors (see eq.\eqref{eqmn4g}), which for convenience is rewritten along with its initial condition as
\begin{equation}
\begin{aligned}
\frac{d }{dt}\vec{S}\left(t\right)=&g\left(t\right)\vec S \left(t\right), &     \vec S \left(0\right) =S_0
\end{aligned}
\end{equation}
whose solution can be expressed as 
\begin{subequations}
\begin{align}
\vec{S}\left(t\right)=&e^{\int_{t_0}^t g\left(t_1\right) dt_1}\vec S \left(0\right)\\
\vec{S}\left(t\right)=&e^{G\left(t_1\right)}\vec S \left(0\right)\label{expsoln}
\end{align}
\end{subequations}
%The approach proposed by Magnus to solve the matrix initial value problem is to express the solution of the exponential of a certain $N\times N$ function $G\left(t,t_0\right)$
%\begin{equation}
%\begin{aligned}
%\vec{S}\left(t\right)=&e^{G\left(t,0\right) }\vec S \left(0\right)
%\end{aligned}
%\end{equation}
%which is subsequently as a series expansion
%\begin{equation}
%\begin{aligned}
%G\left(t,0\right)=&\sum_{k=1}^\infty G_k\left(t,0\right)
%\end{aligned}
%\end{equation}
%writing $G\left(t,0\right)=G\left(t\right)$ for simplicity, the first three series reads thus 
 %\begin{equation}
%\begin{aligned}
%G_1\left(t\right)=&\int_{0}^t g\left(t_1\right) dt_1\\
%G_2\left(t\right)=&\frac{1}{2}\int_{0}^t dt_1 \int_{0}^{t_1} \left[g\left(t_1\right), g\left(t_2\right)\right] dt_2\\
%G_3\left(t\right)=&\frac{1}{6}\int_{0}^t dt_1 \int_{0}^{t_1} dt_2 \int_{0}^{t_2} \Big(\left[g\left(t_1\right),\left[g\left(t_2\right), g\left(t_3\right)\right]\right]+\left [g\left(t_3\right),\left[g\left(t_2\right), g\left(t_1\right)\right]\right] \Big) dt_3
%\end{aligned}
%\end{equation}
%Let us now make an assumption that the matrix $g$ commutes with itself at different times, that is $\left[g\left(t_1\right),g\left(t_2\right)\right]=0$, for simplicity. Of course for better approximation one has to include more terms in the Magnus expansion. With this assumption we will be considering only  the first term only, namely $G_1\left(t\right)$, therefore the solution for eq.\eqref{eqmn4g} 
%\begin{equation}
%\begin{aligned}
%\vec{S}\left(t\right)=&e^{G\left(t_1\right)}\vec S \left(0\right)
%\end{aligned}
%\end{equation}
where $G\left(t_1\right) =\int_{0}^t g\left(t_1\right) dt_1$. Notice here that we are assuming all entries $g_{ij}$, of the matrix $g$, are integrable functions, following which we then define the integral of the matrix as the matrix of the integrals, which mathematically is expressed in the form of $\int g\left(t\right) dt :=(\int g_{ij}\left(t\right) dt)$. The Sylvester's formula is a way of solving analytical function of  a matrix by making use of, only, its eigenvalues \cite{moler2003nineteen,tarantola2006elements}. Therefore given an $N\times N$ matrix, one can readily obtain solution of its exponent by obtaining its eigenvalues and use the Sylvester's formula.  We can write the exponent in eq.\eqref{expsoln} using the formula as 
\begin{equation}
\begin{aligned}
e^{G\left(t_1\right)}=&\sum_{j=1}^{N^2-1}e^{\gamma_j}\prod_{j\neq k=1}^{N^2-1}\frac{G\left(t_1\right)-\gamma_j  I}{\gamma_k-\gamma_j}
\end{aligned}
\label{expSylv}
\end{equation}
where $\gamma_j$ are eigenvalues of $G\left(t_1\right)$, $I$ is an identity matrix, and $e^{\gamma}$ is diagonal matrix. 

At this point it is worth pointing out that one can make use of the Sylvester's formula in two ways. One way, which is also direct use of the formula, is just to plug in the matrix $G$ along with its eigenvalues according to the prescription given by the formula, i.e eq. \eqref{expSylv}. In this case all we need is the knowledge of the eigenvalues of the matrix. The second, and which also needs knowledge of eigenvectors besides eigenvalues, casts the formula into spectral decomposition. Below we will present both approaches and obtain the solution of two level system whose equation of motion is mapped into Bloch equation of motion. We must stress here that for simplicity and practical purpose we assume the matrix elements (of the Hamiltonian) to be integrable either analytically and/or numerically.
%%%%%%%%%%%%%%%%%%%%%%%%%%%%%%%%%%%%%%%%%%%%%%%%%%%%%%%%%%%%%%%%%%%%%%%%%%%%%%%%
\paragraph{\textbf{With Sylvester's Formula :-}}
%%%%%%%%%%%%%%%%%%%%%%%%%%%%%%%%%%%%%%%%%%%%%%%%%%%%%%%%%%%%%%%%%%%%%%%%%%%%%%%%

Let us employ here the Sylvester's formula for two level system discussed earlier, and obtain the analytical solution of the coherence vector, or Bloch vector.  As in the previous case we again assume the system is initially prepared to be on the ground state. We have already seen that the equation of motion for two level system in terms of the $SU(2)$  to be 
\begin{equation}
\begin{aligned}
\frac{d}{dt}\begin{pmatrix}
S_1\\
S_2\\
S_3
\end{pmatrix}=&
 \begin{pmatrix}
  0 & \Delta  & 0 \\
  -\Delta & 0 & -\Omega \\
  0 & \Omega & 0
 \end{pmatrix}
\begin{pmatrix}
S_1\\
S_2\\
S_3
\end{pmatrix}\\
\end{aligned}
\end{equation}
its solution assuming the matrix $g$ is commutable with itself at different times, can be written as
\begin{equation}
\begin{aligned}
\vec{S}\left(t\right)=&e^{G\left(t_1\right)}\vec S \left(0\right)
\end{aligned}
\end{equation}
where $G\left(t_1\right) =\int_{0}^t g\left(t_1\right) dt_1$, and it explicitly is expressible as 
\begin{equation}
\begin{aligned}
G=&\int_{0}^t g\left(t_1\right) dt_1
=\int_{0}^t  dt_1\begin{pmatrix}
  0 & \Delta  & 0 \\
  -\Delta & 0 & -\Omega \\
  0 & \Omega & 0
 \end{pmatrix}\\
 =&\begin{pmatrix}
0 & \int_{0}^{t} \Delta dt_1 & 0 \\
  -\int_{0}^{t} \Delta dt_1 & 0 & -\int_{0}^{t} \Omega dt_1\\
  0 & \int_{0}^{t} \Omega dt_1 & 0
\end{pmatrix}\\
=&\begin{pmatrix}
 0 &  \Delta'  & 0 \\
  -\Delta'  & 0 & -\Omega' \\
  0 &  \Omega'  & 0
 \end{pmatrix}
\end{aligned}
\end{equation}
where $ \Delta' $ and $\Omega'$ are integrated value of $ \Delta$ and $\Omega$ respectively.

The next step is to obtain the eigenvalues, of $g$. As our matrix is low dimensional we are able to find the eigenvalues by solving the characteristics equation $\det (g-\lambda I)=0$ yielding eigenvalues of $\{0, -\sqrt{-\Delta'^2-\Omega'^2} , \sqrt{-\Delta'^2-\Omega'^2} \}$. Then plugging the initial values of the coherence vector, $\vec S\left(0\right)=(0, 0, -1)^T$ which follows from definition of the vector $\vec S=(u_{01},v_{01},w_1)^T$,  we find the solution to be expressible as 
\begin{equation}
\begin{aligned}
S\left(t\right)=~&e^{G\left(t_1\right)}\vec S \left(0\right)=\Big[\sum_{j=1}^{3}e^{\gamma_j}\prod_{j\neq k=1}^{3}\frac{G\left(t_1\right)-\gamma_j  I}{\gamma_k-\gamma_j}\Big]\cdot\vec S \left(0\right)\\
=&\Big[e^{\gamma_1}\frac{(G-\gamma_2I_{3\times3})(G-\gamma_3I_{3\times3})}{(\gamma_1-{\gamma_2})(\gamma_1-\gamma_3)}+e^{\gamma_2}\frac{(G-\gamma_1I_{3\times3})(G-\gamma_3I_{3\times3})}{(\gamma_2-\gamma_1)(\gamma_2-\gamma_3)}+e^{\gamma_3}\frac{(G-\gamma_1I_{3\times3})(G-\gamma_2I_{3\times3})}{(\gamma_3-\gamma_1)(\gamma_3-\gamma_2)}\Big]\cdot\vec S \left(0\right)
\end{aligned}
\label{slvfrm}
\end{equation}
with $\zeta=\sqrt{\Delta'^2+\Omega'^2}$. Recalling the eigenvalues to be $\gamma_1=0$, $\gamma_2=-\imath\zeta$ and $\gamma_3=\imath\zeta$, and upon substituting the corresponding values into eq.\eqref{slvfrm} along with performing some algebra leads the solution to have the following form 
\begin{equation}
\begin{aligned}
\vec{S}\left(t\right)=&
 \begin{pmatrix}
 \frac{\Delta'\Omega'}{\zeta^2}\big(1-\cos\zeta\big)\\
  \frac{\Omega'}{\zeta}\sin\zeta\\
 -\frac{\Delta'^2}{\zeta^2}-\frac{\Omega'^2}{\zeta^2}\cos\zeta
 \end{pmatrix}
\end{aligned}
\label{sylv2l}
\end{equation}
Inspection of solution Eq.\eqref{sylv2l} shows that it has similar form as the one obtained with adiabatic approximation, see Eq.\eqref{soln2lvl}.

%%%%%%%%%%%%%%%%%%%%%%%%%%%%%%%%%%%%%%%%%%%%%%%%%%%%%%%%%%%%%%%%%%%%%%%%%%%%%%%%
\paragraph{\textbf{With Spectral decomposition}}
%%%%%%%%%%%%%%%%%%%%%%%%%%%%%%%%%%%%%%%%%%%%%%%%%%%%%%%%%%%%%%%%%%%%%%%%%%%%%%%%
 We employ here what is known as spectral decomposition method, this method is the factorization of a matrix into its canonical form. In doing so we represent the matrix in terms of its eigenvalues and eigenvectors. It is worth noting that the method is applicable only to diagonalizable matrices, recall that an $n \times n$ matrix is diagonalizable if it has $n$ independent eigenvectors.
\begin{equation}
\begin{aligned}
e^{G\left(t_1\right)}=&\sum P e^{\gamma} P^{-1}
\end{aligned}
\end{equation}
As our matrix $G$, dropping the time argument for clarity, is diagonalizable, it then follows that there exists an invertible matrix $P$ such that $G = PDP^{-1}$, where $D$ is a diagonal matrix of eigenvalues of $G$, and $P$ is a matrix having eigenvectors of $G$ as its columns. In this case, $e^G = Pe^DP^{-1}$.  Where we employed the Taylor series expression of an exponential function along with the relation $(P^{-1} G P)^m = P^{-1} G^m P$. Thus for eigenvectors $\eta_j$, the eigenvalue function is 
\begin{equation}
\begin{aligned}
G\eta_j=&\gamma_j\eta_j
\end{aligned}
\end{equation}
Also using Frobenius covariant where $Q =\eta_j\eta_j^T$ we have, 
\begin{equation}
\begin{aligned}
e^{G}=&\sum_j  e^{\gamma_j}\eta_j\eta_j^T
\end{aligned}
\end{equation}
The Frobenius covariants of our matrix is the projection matrices associated with its eigenvalues and eigenvectors. With which the solution can be expressed as 
\begin{equation}
\begin{aligned}
S\left(t\right)=&\sum_j  e^{\gamma_j}\eta_j\big(\eta_j^T S\left(0\right)\big)
\end{aligned}
\end{equation}
the term $\eta_j^T S\left(0\right)$ is scalar product. 

%%%%%%%%%%%%%%%%%%%%%%%%%%%%%%%%%%%%%%%%%%%%%%%%%%%%%%%%%%%%%%%%%%%%%%%%%%%%%%%%
%\subsection{Sylvester's formula}
%%%%%%%%%%%%%%%%%%%%%%%%%%%%%%%%%%%%%%%%%%%%%%%%%%%%%%%%%%%%%%%%%%%%%%%%%%%%%%%%
To reiterate the Sylvester's formula is a way of solving any exponential function by making use of, only, its eigenvalues. therefore given an $N\times N$ matrix, one can readily obtain solution of its exponent by obtaining its eigenvalues and use the Sylvester's formula. We have expressed the Magnus solution for our coherence vector in terms of the eigenvectors as 
\begin{equation}
\begin{aligned}
S\left(t\right)=&\sum_j  e^{\gamma_j}\eta_j\big(\eta_j^T S\left(0\right)\big)
\end{aligned}
\end{equation}
where $\eta_j\eta_j^T$ being the projection operator. A projection is a linear transformation of matrix $Q$, corresponding to the transformation in an appropriate basis, from a vector space to itself such that $Q^2=Q$. This means whenever $Q$ is applied twice to any vector, it gives the same result as if it were applied once i.e idempotent rule. If we let it to be  $Q_j=\eta_j\eta_j^T$ the solution can be re-expressed as
\begin{equation}
\begin{aligned}
S\left(t\right)=&\sum_j  e^{\gamma_j}Q_j S\left(0\right)
\end{aligned}
\end{equation}
where the projection operator $Q_j$ can be written in Sylvester's formula
\begin{equation}
\begin{aligned}
Q_j=&\prod_{j\neq k=1}^{N^2-1}\frac{G-\gamma_j  I}{\gamma_k-\gamma_j}
\end{aligned}
\end{equation}
where $G\left(t_1\right) dt_1=\int_{0}^t g\left(t_1\right)$, and $\gamma_j$ is the eigenvalues of $G$. It is easy to proof that the Sylvester's formula is indeed a projection operator.

\textbf{Proof}: If $G$ has $n$ distinct eigenvalues, any vector $|\Psi\rangle$ in the $n$ dimensional space can be expanded in terms of the $n$ eigenvectors
\begin{equation}
\begin{aligned}
|\Psi\rangle=&\sum_{k=1}^n \gamma_k |K\rangle
\end{aligned}
\end{equation}
From the eigenvalue equation we have
\begin{equation}
\begin{aligned}
G|j\rangle=& \gamma_k |j\rangle
\end{aligned}
\end{equation}
Therefore to see if $Q_j$, i.e $\prod_{j\neq k=1}^{N^2-1}\frac{G-\gamma_j  I}{\gamma_k-\gamma_j}$, is projection operator, let it act on state $|q\rangle$ as follows
\begin{equation}
\begin{aligned}
%Q_j=&\prod_{j\neq k=1}^{N^2-1}\frac{G-\gamma_j  I}{\gamma_k-\gamma_j}\\
Q_j |q\rangle=&\prod_{j\neq k=1}^{N^2-1}\frac{G-\gamma_j  I}{\gamma_k-\gamma_j} |q\rangle\\
Q_j |q\rangle=&\delta_{jq}|q\rangle
\end{aligned}
\end{equation}
where we used that $G|j\rangle=\gamma_k |j\rangle$
\begin{equation}
\begin{aligned}
 j\neq k, j=q, && \Big(G-\gamma_j\Big)|m\rangle=0|q\rangle\\
 j= k, && \Big(G-\gamma_j\Big)|m\rangle=1|q\rangle
\end{aligned}
\end{equation}
Hence an  operator with eigenvalue $0$ or $1$ is projection operator.

Let us revisit the example discussed in subsection (\ref{adiab}) again and employ the Sylvester's formula for two level system to obtain the analytical solution of the coherence vector, or Bloch vector.  As in the previous case we again assume the system is initially prepared to be on the ground state. To begin with we have already seen that the equation of motion for two level system in terms of the $SU(2)$  to be 
\begin{equation}
\begin{aligned}
\frac{d}{dt}\begin{pmatrix}
S_1\\
S_2\\
S_3
\end{pmatrix}=&
 \begin{pmatrix}
  0 & \Delta  & 0 \\
  -\Delta & 0 & -\Omega \\
  0 & \Omega & 0
 \end{pmatrix}
\begin{pmatrix}
S_1\\
S_2\\
S_3
\end{pmatrix}\\
\end{aligned}
\end{equation}
its solution assuming the matrix $g$ is commutable with itself at different times, can be written as
\begin{equation}
\begin{aligned}
\vec{S}\left(t\right)=&e^{G\left(t_1\right)}\vec S \left(0\right)
\end{aligned}
\end{equation}
where $G\left(t_1\right) =\int_{0}^t g\left(t_1\right) dt_1$. and is expressible as 
\begin{equation}
\begin{aligned}
G=&\int_{0}^t g\left(t_1\right) dt_1
=\int_{0}^t  dt_1\begin{pmatrix}
  0 & \Delta  & 0 \\
  -\Delta & 0 & -\Omega \\
  0 & \Omega & 0
 \end{pmatrix}\\
 =&\begin{pmatrix}
0 & \int_{0}^{t} \Delta dt_1 & 0 \\
  -\int_{0}^{t} \Delta dt_1 & 0 & -\int_{0}^{t} \Omega dt_1\\
  0 & \int_{0}^{t} \Omega dt_1 & 0
\end{pmatrix}\\
=&\begin{pmatrix}
 0 &  \Delta'  & 0 \\
  -\Delta'  & 0 & -\Omega' \\
  0 &  \Omega'  & 0
 \end{pmatrix}
\end{aligned}
\end{equation}
where $ \Delta' $ and $\Omega'$ represent the integrated value of $ \Delta$ and $\Omega$ respectively.

The next step is to obtain the eigensystems, i.e eigenvalues and eigenvectors respectively, of $g$. As our matrix is low dimensional we are able to find the eigenvalue by solving the characteristics equation $\det (g-\lambda I)=0$ yielding eigenvalues of $\{0, -\sqrt{-\Delta'^2-\Omega'^2} , \sqrt{-\Delta'^2-\Omega'^2} \}$ and unnormalized eigenvectors
\begin{equation}
\begin{aligned}
\eta=&
 \begin{pmatrix}
  -\frac{\Omega'}{\Delta'} &\frac{\Delta'}{\Omega'}  & \frac{\Delta'}{\Omega'} \\[0.3em]
  0 & -\frac{\sqrt{-\Delta'^2-\Omega'^2}}{\Omega'} & \frac{\sqrt{-\Delta'^2-\Omega'^2}}{\Omega'} \\[0.3em]
  1 & 1 & 1
 \end{pmatrix}
\end{aligned}
\end{equation}
%\textbf{Check This eigenvector should have been the same as that of the previous $U$}???????????????????
%The inverse of the eigenvector becomes 
%\begin{equation}
%\begin{aligned}
%\eta^{-1}=&
% \begin{pmatrix}
%  -\frac{\Delta'\Omega'}{\Delta'^2+\Omega'^2} & 0  & \frac{\Delta'^2}{\Delta'^2+\Omega'^2 }\\[0.6em]
%  \frac{\Delta'\Omega'}{2\left(\Delta'^2+\Omega'^2\right)}  &  -\frac{\Omega'}{2\left(\sqrt{-\Delta'^2-\Omega'^2}\right)}  &  \frac{\Omega'^2}{2\left(\Delta'^2+\Omega'^2\right)} \\[0.6em]
%  \frac{\Delta'\Omega'}{2\left(\Delta'^2+\Omega'^2\right)} &  \frac{\Omega'}{2\left(\sqrt{-\Delta'^2-\Omega'^2}\right)} & \frac{\Omega'^2}{2\left(\Delta'^2+\Omega'^2\right)} 
% \end{pmatrix}
%\end{aligned}
%\end{equation}
Therefore plugging the initial values of the coherence vector, $\vec S\left(0\right)=(0, 0, -1)^T$ which follows from definition of the vector $\vec S=(u_{01},v_{01},w_1)^T$,  we find the following result for $\eta^{-1}S\left(0\right)$
\begin{equation}
\begin{aligned}
\eta^{-1}S\left(0\right)=-&
 \begin{pmatrix}
  \frac{\Delta'^2}{\Delta'^2+\Omega'^2}\\[0.3em]
  \frac{\Omega'^2}{2\left(\Delta'^2+\Omega'^2\right)}\\[0.3em]
  \frac{\Omega'^2}{2\left(\Delta'^2+\Omega'^2\right)}
 \end{pmatrix}
\end{aligned}
\end{equation}
Therefore our solution is now 
\begin{equation}
\begin{aligned}
S\left(t\right)=&\sum_j  e^{\gamma_j}\eta_j\big(\eta_j^T S\left(0\right)\big)\\
S\left(t\right)=&e^{\gamma_1}\eta_1\big(\eta_1^T S\left(0\right)\big)+e^{\gamma_2}\eta_2\big(\eta_2^T S\left(0\right)\big)+e^{\gamma_3}\eta_3\big(\eta_3^T S\left(0\right)\big)
\end{aligned}
\end{equation}
with $\zeta=\sqrt{\Delta'^2+\Omega'^2}$ we note that $\gamma_1=0$, $\gamma_2=-\imath\zeta$ and $\gamma_3=\imath\zeta$, we then obtain
\begin{equation}
\begin{aligned}
S\left(t\right)=&e^0\begin{pmatrix}
 -\frac{\Omega'}{\Delta'}\\
 0\\
 1
\end{pmatrix}\Big(\frac{\Delta'^2}{\zeta^2} \Big) +
e^{-\imath\zeta}\begin{pmatrix}
 -\frac{\Delta'}{\Omega'}\\
 -\frac{-\imath\zeta}{\Omega'}\\
 1
\end{pmatrix}\Big(\frac{\Omega'^2}{2\zeta^2} \Big) +
e^{\imath\zeta}\begin{pmatrix}
 \frac{\Delta'}{\Omega'}\\
 \frac{-\imath\zeta}{\Omega'}\\
 1
\end{pmatrix}\Big(-\frac{\Omega'^2}{2\zeta^2} \Big)
\end{aligned}
\end{equation}
which after some algebra the solution becomes 
\begin{equation}
\begin{aligned}
\vec{S}\left(t\right)=&
 \begin{pmatrix}
 \frac{\Delta'\Omega'}{\zeta^2}\big(1-\cos\zeta\big)\\
  \frac{\Omega'}{\zeta}\sin\zeta\\
 -\frac{\Delta'^2}{\zeta^2}-\frac{\Omega'^2}{\zeta^2}\cos\zeta
 \end{pmatrix}
\end{aligned}
\label{sylv2l}
\end{equation}
Inspection of solution Eq.\eqref{sylv2l} shows that it has similar form as the one obtained with adiabatic approximation, see Eq.\eqref{soln2lvl}.

%%%%%%%%%%%%%%%%%%%%%%%%%%%%%%%%%%%%%%%%%%%%%%%%%%%%%%%%%%%%%%%%%%%%%%%%%%%%%%%%
\subsubsection{Sylvester's formula for degenerate eigenvalues}
%%%%%%%%%%%%%%%%%%%%%%%%%%%%%%%%%%%%%%%%%%%%%%%%%%%%%%%%%%%%%%%%%%%%%%%%%%%%%%%%
So far we have dealt with Sylvester's formula for system with distinct eigenvalue. In this section we look at system with degenerate eigenvalues. For clarity we define degenerate eigenvalue as an eigenvalue $\gamma$ that corresponds to two or more different linearly independent eigenvectors. Mathematically this statement can be expressed as $ GV_{1}=\gamma V_{1}$  and $ GV_{2}=\gamma V_{2}$, where $ V_{1}$ and $V_{2}$ are linearly independent eigenvectors. The eigenvalues, in the context of this paper, give the measurable values of physical observables whereas the corresponding eigenstates give the possible states in which the system may be found.% The measurable values of the energy of a quantum system are given by the eigenvalues of the Hamiltonian operator, while its eigenstates give the possible energy states of the system.

%Therefore we can now express our solution for the coherence vector in terms of the Sylvester's formula as, iff we have distinct eigenvalues 
%\begin{equation}
%\begin{aligned}
%S\left(t\right)=&\sum_j  e^{\gamma_j}\prod_{j\neq k=1}^{N^2-1}\frac{G-\gamma_j  I}{\gamma_k-\gamma_j} S\left(0\right)
%\end{aligned}
%\end{equation}
In such cases the preceding formula is not readily applicable to get the solution, this implies that if we have degenerate eigenvalues Sylvester's formula needs to be modified in order for us to use it to obtain the required solution. To this end let $m_j$ denote the multiplicity of the the eigenvalue $\gamma_j$, the exponent in solution $S\left(t\right)=e^{G} S\left(0\right)$ takes the form \cite{tarantola2006elements}
\begin{equation}
\begin{aligned}
%S\left(t\right)=&e^{G} S\left(0\right)\\
e^{G} =&\sum_j\Big[\sum_{k=0} ^{m_j-1} b_k\left(\gamma_j\right)\Big(G-\gamma_j  I\Big)\Big]\prod_{j\neq i=1}\big(G-\gamma_i  I\big)^{m_j}
\end{aligned}
\label{sylvtrdegen}
\end{equation}
where the sum is performed over all distinct eigenvalue $\gamma_j$, and where the $b_k\left(\gamma_j\right)$ are the scalars
\begin{equation}
\begin{aligned}
b_k\left(\gamma_j\right)=&\frac{1}{k!}\frac{d^k}{d\gamma^k}\Big[\frac{e^{\gamma}}{\prod_{j\neq i=1}^{N^2-1}\big(\gamma-\gamma_j  I\big)^{m_j}}\Big]_{\gamma=\gamma_i}
\end{aligned}
\end{equation}

\textbf{Example - ii}: As an instance let us now consider a three level-$\Lambda$ system of states $\{|0\rangle, |1\rangle,|2\rangle\}$. While there is no coupling between states $|0\rangle\leftrightarrow|2\rangle$, levels $|0\rangle\leftrightarrow|1\rangle$ are coupled by a coherent light $\alpha\left(t\right)=\frac{\hbar}{2}\Omega_p\left(t\right)$ whereas the levels $|1\rangle\leftrightarrow|2\rangle$ are coupled by another coherent light $\beta\left(t\right)=\frac{\hbar}{2}\Omega_s\left(t\right)$, with $\alpha\left(t\right), \beta\left(t\right)$ respectively defined to be the half Rabi frequency of pump and Stokes light. The Hamiltonian of such system is well studied and has three distinct eigenvalues \cite{Shore:2008aa, fewell1997coherent, vitanovst99}. But as our target is to seek analytical solution via Sylvester's formula we need to map the Hamiltonian into bigger space, one way to achieve this is by using Lie algebraic approach as outlined in \cite{2018JPhCS.987a2033H}. In this approach the physical observables are expressed in terms of generators of a Lie group. The Hamiltonian and density matrix are expressible as a linear combination of these generator. It shown that the Hamiltonian in this bigger space,  $g_{\alpha\beta}$,  in a matrix form to be
\begin{equation}
\begin{aligned}g=&
\begin{pmatrix}
0 & 0 & 0 & \Delta & 0 & \beta & 0 & 0\\
0 & 0 & 0 & 0 & -\Delta & -\alpha & 0 & 0\\
0 & 0 & 0 & \beta & -\alpha & 0 & 0 & 0\\
-\Delta & 0 & -\beta & 0 & 0 & 0 & 2\alpha & 0\\
0 & \Delta & \alpha & 0 & 0 & 0 & -\beta & \sqrt{3}\beta\\
-\beta & \alpha & 0 & 0 & 0 & 0 & 0 & 0\\
0 & 0 & 0 & -2\alpha & \beta & 0 & 0 & 0\\
0 & 0 & 0 & 0 & -\sqrt{3}\beta & 0 & 0 & 0
\end{pmatrix}
\end{aligned}
\end{equation}
The eigenvalues, $\gamma_j$, of this Hamiltonian is thus obtained to be as follows. 
\begin{equation}
\begin{aligned}
\gamma_{1,2}=&0,\\
\gamma_{3,4}=&\pm\sqrt{-4\alpha^2-4\beta^2-\Delta^2}\\
\gamma_{5,6,7,8}=&\pm\frac{\sqrt{-2\alpha^2-2\beta^2-\Delta^2\mp\Delta\sqrt{4\alpha^2+4\beta^2-\Delta^2}}}{\sqrt{2}}\\
%\gamma_{7,8}=&\pm\frac{\sqrt{-2\alpha^2-2\beta^2-\Delta^2+\Delta\sqrt{4\alpha^2+4\beta^2-\Delta^2}}}{\sqrt{2}}\\
\end{aligned}
\end{equation}
As we can see that two of the eigenvalues are zero, say  $\gamma_1=\gamma_2=0$, and the remaining six eigenvalues are pure imaginary. Therefore because we have two degenerate eigenvalues we have to use the second form of the Sylvester's formula, which is eq.\eqref{sylvtrdegen}, note we here take matrix $G$ to be the matrix after performing integration to the matrix, i.e. its elements. %namely $G_2\left(t\right)$\\
Assuming the system is initially prepared to be on the ground state the initial condition of the vector is taken to be $S\left(0\right)=(0,0,0,0,0,0,-1,-\frac{1}{\sqrt{3}})^T$, the evolution of the eight dimensional coherence vector is then obtained, using Sylvester's formula, as follows 
\begin{equation}
\begin{aligned}
S\left(t\right)=&e^{G} S\left(0\right)\\
=&\Big[\Big(b_0\left(\gamma_1=0\right)+b_1\left(\gamma_2=0\right)\big(G-\gamma(=0)  I_{8\times8}\big)\Big)\prod_{k\neq 1,2 k=3}^8\big(G-\gamma_k  I_{8\times8}\big)
+\sum_{j=3}^8 e^{\gamma_j}\prod_{j\neq k=1}^{8}\frac{G-\gamma_k  I_{8\times8}}{\gamma_j-\gamma_k}\Big]S\left(0\right)
\end{aligned}
\end{equation}
where $I_{8\times8}$ is an $8\times8$ identity matrix and the scalar elements $b_n$ ( for $n=0,1$) corresponding to the two degenerate eigenvalues (with $\gamma_{1,2}=0$) are expressed as
\begin{equation}
\begin{aligned}
b_0\left(\gamma_i\right)=&\frac{1}{0!}\frac{d^0}{d\gamma^0}\Big[\frac{e^{\gamma}}{\prod_{k=3}^{8}\big(\gamma-\gamma_k  I\big)}\Big]_{\gamma=0}
=\frac{1}{\prod_{k=3}^{8}\big(-\gamma_k  \big)  }
\end{aligned}
\end{equation}
and 
\begin{equation}
\begin{aligned}
b_1\left(\gamma_i\right)=&\frac{1}{1!}\frac{d}{d\gamma}\Big[\frac{e^{\gamma}}{\prod_{k=3}^{8}\big(\gamma-\gamma_k  I\big)}\Big]_{\gamma=0}\\
%=&\frac{\big(\frac{d }{d\gamma}e^{\gamma}\big)\prod_{k=3}^{8}\big(\gamma-\gamma_k  I\big)-e^{\gamma}\big(\frac{d}{d\gamma}\prod_{k=3}^{8}\big(\gamma-\gamma_k  I\big)\big)}{\Big[\prod_{k=3}^{8}\big(\gamma-\gamma_k  I\big)\Big]^2}\Big|_{\gamma=0}\\
=&\frac{\prod_{k=3}^{8}\big(-\gamma_k  I\big)-\prod_{k=3}^{8}\big(1-\gamma_k  I\big)}{\Big[\prod_{k=3}^{8}\big(-\gamma_k  I\big)\Big]^2}\\
\end{aligned}
\end{equation}

%%%%%%%%%%%%%%%%%%%%%%%%%%%%%%%%%%%%%%%%%%%%%%%%%%%%%%%%%%%%%%%%%%%%%%%%%%%%%%%%%
%%%%%%%%%%%%%%%%%%%%%%%%%%%%%%%%%%%%%%%%%%%%%%%%%%%%%%%%%%%%%%%%%%%%%%%%%%%%%%%%%
%\makeatletter
%\renewcommand{\thesection}{\thechapter. \@arabic\c@section}
%\renewcommand{\thesubsection}{\thesection. \@arabic\c@section}
%\renewcommand{\thesubsubsection}{\thesubsection. \@arabic\c@section}
%\makeatother
%\begin{appendices}

%%%%%%%%%%%%%%%%%%%%%%%%%%%%%%%%%%%%%%%%%%%%%%%%%%%%%%%%%%%%%%%%%%%
%Steps for typesetting document with bibtex
%1) Latex
%2) Bibtex
%3) Latex
%4) Latex
\bibliographystyle{plain}
\bibliography{References}
\end{document}